# Optical force on toroidal nanostructures: toroidal dipole versus renormalized electric dipole

Xu-Lin Zhang,[1,2] S. B. Wang,[1] Zhifang Lin,[3,4] Hong-Bo Sun,[2] and C. T. Chan[1,*]

[1]*Department of Physics and Institute for Advanced Study, The Hong Kong University of Science and Technology, Clear Water Bay, Hong Kong, China*
[2]*State Key Laboratory on Integrated Optoelectronics, College of Electronic Science and Engineering, Jilin University, Changchun, China*
[3]*State Key Laboratory of Surface Physics and Key Laboratory of Micro and Nano Photonics Structure (Ministry of Education), Fudan University, Shanghai 200433, China*
[4]*Collaborative Innovation Center of Advanced Microstructures, Fudan University, Shanghai 200433, China*

[*]Corresponding author: phchan@ust.hk

We study the optical forces acting on toroidal nanostructures. A great enhancement of optical force is unambiguously identified as originating from the toroidal dipole resonance based on the source-representation, where the distribution of the induced charges and currents is characterized by the three families of electric, magnetic, and toroidal multipoles. On the other hand, the resonant optical force can also be completely attributed to an electric dipole resonance in the alternative field-representation, where the electromagnetic fields in the source-free region are expressed by two sets of electric and magnetic multipole fields based on symmetry. The confusion is resolved by conceptually introducing the irreducible electric dipole, toroidal dipole, and renormalized electric dipole. We demonstrate that the optical force is a powerful tool to identify toroidal response even when its scattering intensity is dwarfed by the conventional electric and magnetic multipoles.

The notions of toroidal multipoles have been widely used in different disciplines of physics [1-7]. The simplest model for a toroidal dipole can be visualized as current flowing on the surface of a torus along its meridians. Such complex current distributions are difficult to excite but thanks to recent advances in material fabrication [8-10], the toroidal metamaterials were theoretically predicted [11] and experimentally demonstrated at microwave frequencies [12], followed by various studies on their intriguing properties [13-19]. The need for introducing toroidal multipoles is due to the fact that a general type of the charge/current distribution, either free or induced, cannot be represented exclusively by the standard electric and magnetic multipoles as defined in textbooks. From the mathematical point of view, the current is a vector with three independent components, while the charge is a scalar. With the continuity equation, one needs three functionals to characterize a general type of source, corresponding to three families of multipoles. This constitutes the source-representation, so termed as it focuses on parametrizing the source, either free or induced charge/current. On the other hand, a divergenceless vector function can be expanded in terms of two sets of multipole fields, termed electric and magnetic multipole fields described by the vector spherical functions [20], with the expansion coefficients designated as electric and magnetic multipole coefficients (see, e.g., [21]). This forms the field-representation since it focuses on the radiated fields. The electromagnetic fields in the source-free region can be defined in the field-representation based on electric and magnetic multipoles, without resorting to the third type of toroidal multipoles [20-23]. This naturally leads to a question as to whether toroidal excitation is indispensable to characterize the electromagnetic fields. This issue has been largely ignored since the toroidal excitation is typically obscured by much stronger electric or magnetic multipole excitations. But with more and more "toroidal excitations" being reported in the literature, it is highly desirable that we can reconcile the source-representation and the field-representation.

We clarify this issue by studying the optical forces acting on nanostructures that support toroidal excitations. On one hand, the optical force can produce a measurable physical consequence that has been extensively employed to manipulate particles [24-32]. On the other hand, it can be theoretically calculated either, in the source-representation, by the interaction between the external fields and the *three* families of multipoles induced on the object or, in the field-representation, through the coupling between the incident fields and the *two* types of multipoles excited on the scatter. Hence, it serves as a good platform to address the aforementioned confusion. While the toroidal dipole is necessary for the description of optical forces in the source-representation and may overwhelmingly dominate the optical force when toroidal dipole resonance is excited, its role is entirely attributed to the renormalized electric dipole resonance in the field-representation. More importantly, we will show that optical force can serve as a definitive quantity for the detection of toroidal response in nanostructures even when its scattering intensity is overwhelmed by the conventional electric and magnetic multipoles.

We start by considering an object illuminated by a time-harmonic incident wave. The time-averaged optical force can be written as an integral over a closed spherical surface $S_\infty$ with radius $R_S \to \infty$ [27,30],

$$\left\langle \mathbf{F}^{\text{exact}} \right\rangle = -\frac{1}{c} \oint_{S_\infty} \left[ \left\langle \mathbf{S}_{\text{mix}} \right\rangle + \left\langle \mathbf{S}_s \right\rangle \right] d\sigma, \quad (1)$$

where $c$ is the speed of light in vacuum, $\left\langle \mathbf{S}_{\text{mix}} \right\rangle = 1/2 \text{Re}\left[ \mathbf{E}_i \times \mathbf{H}_s^* + \mathbf{E}_s \times \mathbf{H}_i^* \right]$ and $\left\langle \mathbf{S}_s \right\rangle = 1/2 \text{Re}\left[ \mathbf{E}_s \times \mathbf{H}_s^* \right]$, with $\mathbf{E}_i$ ($\mathbf{H}_i$) and $\mathbf{E}_s$ ($\mathbf{H}_s$) denoting, respectively, the incident and scattered electric (magnetic) fields. Equation (1) is formally "exact" provided that the scattered fields are known.

In the source-representation, the scattered field from the object is written in terms of the vector potential depending on the induced current density $\mathbf{J}(\mathbf{r}')e^{i\omega R/c}$ on the object, $\mathbf{A}(\mathbf{r}) = \mu_0 / (4\pi) \int_V \left[ \mathbf{J}(\mathbf{r}')e^{i\omega R/c} / R \right] dv'$, where $\omega$ is the angular frequency, $\mu_0$ is the free space permeability, $R = |\mathbf{r} - \mathbf{r}'|$, and the time dependence $e^{-i\omega t}$ has been assumed and suppressed. The vector potential can be Taylor-expanded into the primitive multipole terms, with the first few moments expressed as [22]

$$p_i = -\frac{1}{i\omega} \int_V J_i dv', \quad m_i = \frac{1}{2} \int_V (\mathbf{r}' \times \mathbf{J})_i dv',$$
$$q_{ij}^{(e)} = -\frac{1}{i\omega} \int_V (r_i' J_j + J_i r_j') dv', \quad q_{ij}^{(m)} = \frac{2}{3} \int_V (\mathbf{r}' \times \mathbf{J})_i r_j' dv', \quad (2)$$

which are the Cartesian components of the *primitive* electric dipole moment (**p**), magnetic dipole moment (**m**), electric quadrupole moment (**q**[(e)]), and magnetic quadrupole moment (**q**[(m)]), respectively. These primitive multipoles are actually not widely used because they are non-traceless and asymmetric. One usually extracts the irreducible parts from the primitive multipoles, and attributes the remaining terms to the toroidal multipoles, leading to the irreducible electric and magnetic multipoles, with a few lowest order moments given by $p_i$ and $m_i$ in Eq. (2), and

$$Q_{ij}^{(e)} = -\frac{1}{i\omega}\int_V\left(r_i'J_j + J_ir_j' - \frac{2}{3}(\mathbf{r}'\cdot\mathbf{J})\delta_{ij}\right)dv',$$

$$Q_{ij}^{(m)} = \frac{1}{3}\int_V\left[(\mathbf{r}'\times\mathbf{J})_ir_j' + (\mathbf{r}'\times\mathbf{J})_jr_i'\right]dv', \quad (3)$$

$$t_i = \frac{1}{10}\int_V\left[(\mathbf{r}'\cdot\mathbf{J})\mathbf{r}' - 2(\mathbf{r}'\cdot\mathbf{r}')\mathbf{J}\right]_i dv',$$

where $Q_{ij}^{(e)}$, $Q_{ij}^{(m)}$, and $t_i$ are the Cartesian components of the *irreducible* electric quadrupole moment, magnetic quadrupole moment, and the toroidal dipole moment, respectively. We note that the irreducible electric and magnetic dipoles and the primitive ones are the same but they are different for higher order moments such as quadrupole moments (e.g., $Q_{ij}^{(e)}$ versus $q_{ij}^{(e)}$).

With the multipole moments determined by the induced current density, Eq. (1) can be written as a sum [27,30] of the incident forces $\mathbf{F^p}$, $\mathbf{F^m}$, and $\mathbf{F^t}$, due to the direct interaction between the incident fields and the irreducible electric, magnetic and toroidal dipoles (**p**, **m**, **t**), and the recoil force $\mathbf{F}^{\text{int}}$ [30], resulting from dipole interference,

$$\mathbf{F^p} = \frac{1}{2}\text{Re}\left[\left(\nabla\mathbf{E}_i^*\right)\cdot\mathbf{p}\right],$$

$$\mathbf{F^m} = \frac{1}{2}\text{Re}\left[\left(\nabla\mathbf{B}_i^*\right)\cdot\mathbf{m}\right],$$

$$\mathbf{F^t} = -\frac{k}{2c}\text{Im}\left[\left(\nabla\mathbf{E}_i^*\right)\cdot\mathbf{t}\right], \quad (4)$$

$$\mathbf{F}^{\text{int}} = -\frac{k^4}{12\pi\varepsilon_0 c}\text{Re}\left[\mathbf{p}\times\mathbf{m}^*\right] + \frac{k^5}{12\pi\varepsilon_0 c^2}\text{Im}\left[\mathbf{m}\times\mathbf{t}^*\right],$$

where $k$ and $\varepsilon_0$ are, respectively, the wave vector and the free space permittivity. Here, we are only concerned with the optical force in the *k*-direction and we focus only on the dipole terms in Eq. (4) because they dominate the force acting on a sub-wavelength nanostructure.

In the field-representation, one starts by expanding the scattered field directly in terms of the outgoing and divergenceless vector spherical functions $\mathbf{N}_{mn}^{(3)}(k,\mathbf{r})$ and

$\mathbf{M}_{mn}^{(3)}(k, \mathbf{r})$ as [20,23,33,34] $\mathbf{E}_s = \sum_{n,m} iE_{mn} \left[ a_{mn} \mathbf{N}_{mn}^{(3)}(k, \mathbf{r}) + b_{mn} \mathbf{M}_{mn}^{(3)}(k, \mathbf{r}) \right]$. Here $a_{mn}$ and $b_{mn}$ are the scattering coefficients (also known as the electric and magnetic multipole coefficients [21], apart from the different normalization factors) that can be determined by doing a field projection [21,23,35] and $E_{mn} = |E_0| i^n \sqrt{(2n+1)(n-m)!} / \sqrt{n(n+1)(n+m)!}$ with $|E_0|$ characterizing the intensity of the incident electric field. For $n = 1$, the coefficients $a_{mn}$ and $b_{mn}$ are directly related to the electric and magnetic dipole moments. The dipoles so defined distinguish themselves from the irreducible dipoles defined in Eq. (2), as can be readily perceived from the exclusion of the toroidal dipole in the field-representation. They are therefore termed "renormalized dipoles".

Inserting the scattered fields into Eq. (1), we can similarly decompose the optical force into incident forces $\mathbf{F}^{\bar{\mathbf{p}}}$ and $\mathbf{F}^{\bar{\mathbf{m}}}$, arising from the direct interaction between incident field and the renormalized dipoles, and the recoil force $\mathbf{F}^{\overline{\text{int}}}$, due to the interference of the renormalized dipoles, with its $k$- ($z$-) component written as [32,35]

$$F_z^{\bar{\mathbf{p}}} = \frac{2\pi\varepsilon_0}{k^2} |E_0|^2 \sum_{n=1,m} \text{Re}\left[ g_{mn} a_{m,n} p_{m,n+1}^* + l_{mn} q_{m,n} a_{m,n}^* \right],$$

$$F_z^{\bar{\mathbf{m}}} = \frac{2\pi\varepsilon_0}{k^2} |E_0|^2 \sum_{n=1,m} \text{Re}\left[ g_{mn} b_{m,n} q_{m,n+1}^* + l_{mn} p_{m,n} b_{m,n}^* \right], \quad (5)$$

$$F_z^{\overline{\text{int}}} = -\frac{4\pi\varepsilon_0}{k^2} |E_0|^2 \sum_{n=1,m} \text{Re}\left( l_{mn} a_{m,n} b_{m,n}^* \right),$$

where $p_{m,n}$ and $q_{m,n}$ are the expansion coefficients for the incident wave in terms of vector spherical functions, and $g_{mn}$ and $l_{mn}$ are given in the Supplemental material [35]. Here we use overlines to denote the renormalized dipoles (i.e., $\bar{\mathbf{p}}$, $\bar{\mathbf{m}}$), in comparison with the irreducible dipoles and the toroidal dipole (i.e., $\mathbf{p}$, $\mathbf{m}$, $\mathbf{t}$).

We therefore see that the optical force acting on sub-wavelength nanostructures can be written as a summation of forces acting on the toroidal dipole as well as the irreducible electric and magnetic dipoles, or alternatively, interpreted as a summation of forces exerting exclusively on the renormalized electric and magnetic dipoles. We will identify the relationship between the irreducible and renormalized dipoles by performing a numerical study on toroidal nanostructures.

We consider a toroidal nanostructure shown in Fig. 1(a) which consists of four gold helixes arranged head-to-tail along a loop. Each helix individually supports a magnetic dipole resonance in the optical regime as shown in the Supplemental material [35], and they will collectively support a toroidal dipole resonance. The incident wave is a linearly *x*-polarized plane wave propagating in the *z*-direction. In the optical regime, the permittivity of gold is described by the Drude model with plasma frequency $\omega_p = 2\pi \times 2.175 \times 10^{15}$ rad/s and damping frequency $\omega_c = 2\pi \times 6.5 \times 10^{12}$ rad/s [36]. We first use the source-representation, in which case the response of the nanostructure can be described by the irreducible dipoles and the toroidal dipole. The dipole moments can be calculated (Eqs. (2) and (3)) using the current density obtained with a finite-element method package COMSOL [37]. The calculated scattering cross sections (see Supplemental material) for these induced dipole moments are shown in Fig. 1(c). Only those components that contribute to the incident force are shown here, that is, $p_x$, $m_y$, and $t_x$. Two noticeable resonant peaks (at ~64 THz and ~71 THz) are identified, which originate from $m_y$ and $t_x$ resonances, respectively. The current density distribution for the resonance at ~71 THz is illustrated in Fig. 1(b), where the threading of the magnetic field lines (black lines) around the nanostructure confirms the excitation of a toroidal dipole resonance [12]. We show the *z*-component of the optical forces (Eq. (4)) in Fig. 1(d), with two peaks at the frequencies coinciding with those of the scattering cross sections in Fig. 1(c). The resonant force at ~64 THz can be attributed to the irreducible magnetic dipole moment $m_y$ while the one at ~71 THz can mainly be associated with the toroidal dipole moment $t_x$. This provides good evidence that the toroidal dipole is indispensable based on the source-representation. The total force acting on all these dipoles ($F_z^{\mathbf{p}} + F_z^{\mathbf{m}} + F_z^{\mathbf{t}} + F_z^{\text{int}}$) could well reproduce the exact result ($F_z^{\text{exact}}$) that is calculated by Eq. (1) using numerically obtained scattered fields [37], confirming the validity of our approach. The toroidal dipole induced force in other similar nanostructures (e.g., changing the handedness of the helixes) are also studied and can be found in the Supplemental material [35], which can further verify the important role of the toroidal moment in understanding the physics of the optical forces.

We now use the field-representation. We first calculate the partial scattering cross sections (see Supplemental material) from each renormalized dipoles and the results are shown in Fig. 2(a). We note that at ~64 THz and ~71 THz, resonances of a renormalized magnetic dipole $\bar{m}_y$ and a renormalized electric dipole $\bar{p}_x$ are identified respectively. We then calculate the optical forces using Eq. (5) with the coefficients $a_{mn}$, $b_{mn}$, $p_{mn}$, and $q_{mn}$ evaluated by field projections [35]. Figure 2(b) shows that the strong enhancement of force at the resonance of ~71 THz is contributed by $F_z^{\bar{\mathbf{p}}}$, coming from the interaction between the incident field and the renormalized electric dipole moment, while the same resonance force in the source-representation has been interpreted as a summation of the contributions from the irreducible electric dipole and the toroidal dipole, with major contribution coming from the toroidal dipole in Fig. 1(d). The resonance at ~64 THz in Fig. 2(b) can be attributed to the renormalized magnetic dipole, in agreement with the interpretation in Fig. 1(d). By comparing the force contributions between the two representations (see Figs. 1(d) and 2(b)), we can infer that $\mathbf{F}^{\mathbf{p}} + \mathbf{F}^{\mathbf{t}} \approx \mathbf{F}^{\bar{\mathbf{p}}}$ and $\mathbf{F}^{\mathbf{m}} \approx \mathbf{F}^{\bar{\mathbf{m}}}$. Actually, the electric far-field radiation from the irreducible electric dipole and the toroidal dipole can be written as $\mathbf{n} \times (\mathbf{p} \times \mathbf{n})$ and $(ik/c)\mathbf{n} \times (\mathbf{t} \times \mathbf{n})$ respectively, multiplied by a same factor $k^2 e^{ikr}/(4\pi\varepsilon_0 r)$, with $\mathbf{n}$ denoting the radial direction [30]. At the far-field, one cannot tell whether the radiation field comes from an irreducible electric dipole or a toroidal dipole, but only find a renormalized electric dipole instead. We also note that the similarity of the far field radiation pattern of the irreducible electric dipole and the toroidal dipole was very recently exploited to produce 'anapoles' [38].

In general, toroidal resonances are weak and their contributions to the scattering power are usually obscured by the much stronger irreducible dipoles. In this case, optical force can serve to detect the toroidal response. To illustrate this point, we plot in Fig. 3 the sum of the scattering cross sections due to all dipolar components (i.e. $x$, $y$, $z$) of the nanostructure shown in Fig. 1(a). The results based on the source-representation agree well with those based on the field-representation. While the effect of the toroidal dipole on the total scattering cross section is barely noticeable (the small peak of $C_{\text{sca}}^{\mathbf{p+t}}$ or $C_{\text{sca}}^{\bar{\mathbf{p}}}$ at ~71 THz), it gives the dominant contribution to the optical force at ~71 THz (see Fig. 1(d)). The strong $C_{\text{sca}}^{\mathbf{p+t}}$ (or $C_{\text{sca}}^{\bar{\mathbf{p}}}$) here is due to the $z$-component of the corresponding electric dipole, which yields a negligible optical force. The reason that optical force can reveal the effect of toroidal dipole is due to the underlying symmetry information incorporated in the force formula. The scattering cross section only gives the information of the total number of photons scattered, while optical force has the extra dependence on the distribution of the scattered photons. Therefore in cases where the scattering cross section of the toroidal dipole is buried in the background and hardly detectable, the toroidal response may still manifest itself through optical force due to its selectivity in symmetry.

We note that our theory can also be applied to other toroidal structures [12,13]. We have carried out similar studies in the Supplemental material [35] for these structures and we reach the same conclusion. The theory also applies to the microwave regime [35].

To summarize, we have studied the optical forces acting on toroidal nanostructures based on both the source-representation and the field-representation. Some confusion in the understanding of various multipoles has been clarified by introducing and distinguishing the primitive multipoles, irreducible multipoles, and renormalized multipoles. A resonant optical force on the toroidal nanostructure can be understood using the source-representation as a result of the incident wave interacting with the induced toroidal dipole. Or, using the field-representation, it is simply interpreted as due to the interaction between the incident field and the renormalized electric dipole. Each viewpoint has its own advantage. The field-representation expedites the solution of optical scattering problem in many cases, but the source-representation appears more advantageous to understand the physics of the optical force. We also show that optical force enables the observation of the toroidal response of a nanostructure even when its effect on scattering power is overwhelmed by the conventional multipoles. It may also find applications in sorting/selecting nanoparticles with toroidal response.

### ACKNOWLEDGMENS
This work is supported by Hong Kong RGC through AOE/P-02/12 and CUHK1/CRF/12G-1. Z. L. is supported by NNSFC through Grant No. 11174059 and MOE of China (B06011). We thank Profs. Z. Q. Zhang and J. Ng for their valuable comments and suggestions.

**Figures and their captions**

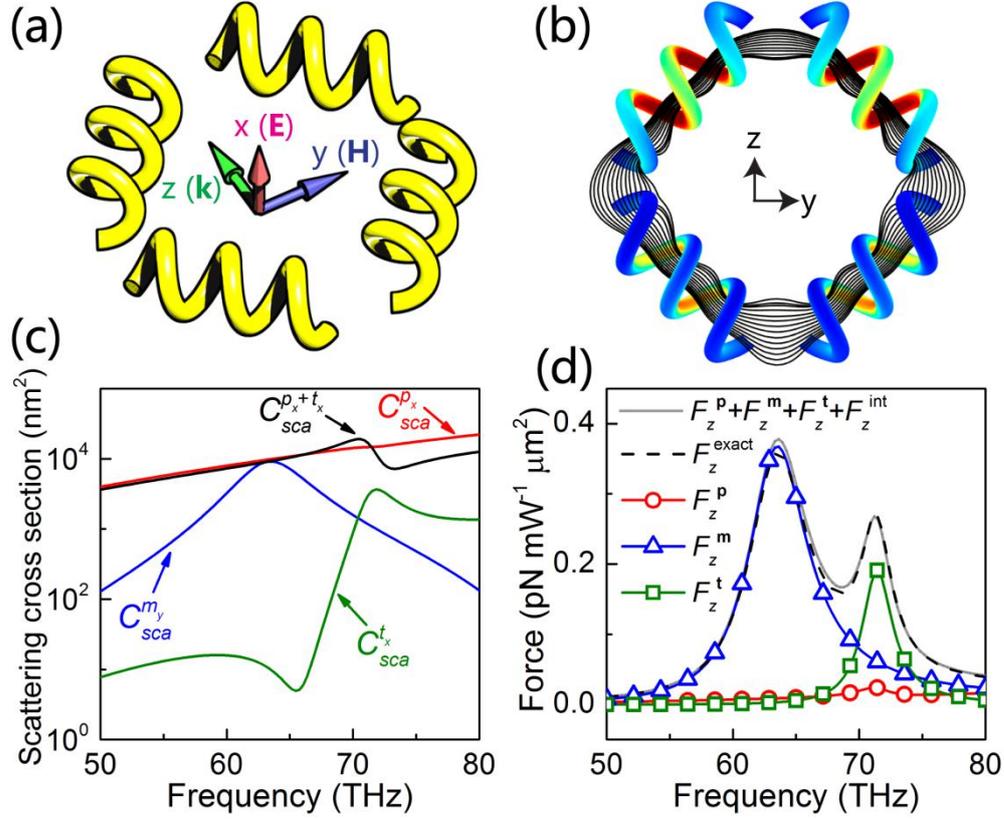

FIG. 1. (Color online) (a) Schematic diagram of the designed nanostructure that supports a toroidal dipole resonance. Each helix has an inner radius of 30 nm, outer radius of 100 nm, and a pitch of 200 nm. (b) The intensity distribution of the current density associated with magnetic field lines (black lines) for the toroidal dipole resonance at ~71 THz. (c) Calculated scattering cross sections and (d) $z$-component of optical forces based on the source-representation. The toroidal response manifests itself through a resonant peak in the total optical force at ~71 THz.

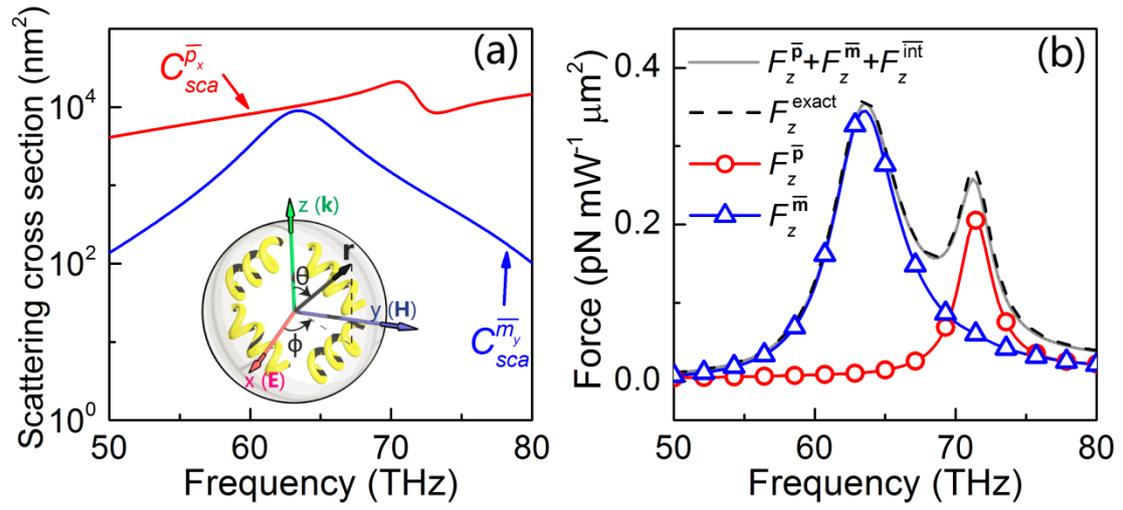

FIG. 2. (Color online) (a) Calculated scattering cross sections and (b) $z$-component of optical forces acting on the renormalized dipoles for the toroidal nanostructure, according to the field-representation.

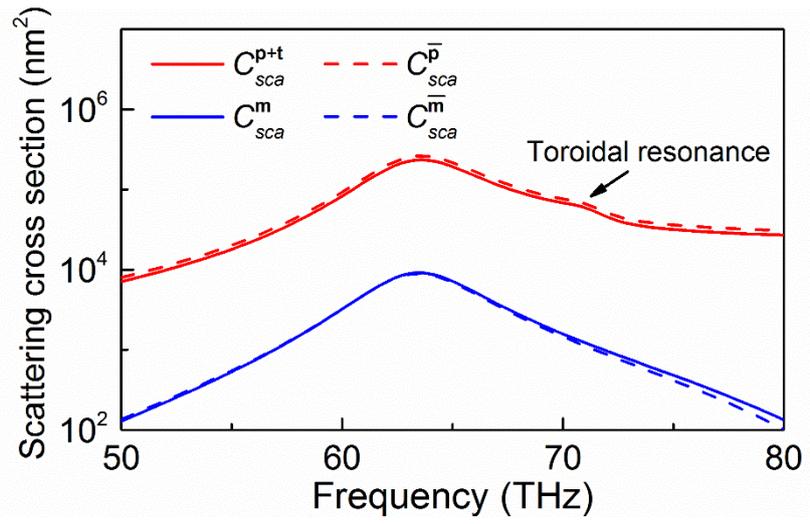

FIG. 3. (Color online) Calculated scattering cross sections due to all dipolar components based on source- (solid lines) and field-representation (dashed lines).

# Supplementary material for "Optical force on toroidal nanostructures: toroidal dipole versus renormalized electric dipole"

Xu-Lin Zhang,[1,2] S. B. Wang,[1] Zhifang Lin,[3,4] Hong-Bo Sun,[2] and C. T. Chan[1,*]

[1]*Department of Physics and Institute for Advanced Study, The Hong Kong University of Science and Technology, Clear Water Bay, Hong Kong, China*
[2]*State Key Laboratory on Integrated Optoelectronics, College of Electronic Science and Engineering, Jilin University, Changchun, China*
[3]*State Key Laboratory of Surface Physics and Key Laboratory of Micro and Nano Photonics Structure (Ministry of Education), Fudan University, Shanghai 200433, China*
[4]*Collaborative Innovation Center of Advanced Microstructures, Fudan University, Shanghai 200433, China*

[*]Corresponding author: phchan@ust.hk

**I Calculation of the scattering cross section**

**(1) For irreducible dipoles and toroidal dipole**:

Based on the source-representation, the radiated powers *I* from the irreducible dipoles and the toroidal dipole can be written as [1,2]

$$I = \frac{1}{4\pi\varepsilon_0}\left(\frac{2\omega^4}{3c^3}|\mathbf{p}|^2 + \frac{2\omega^4}{3c^5}|\mathbf{m}|^2 + \frac{4\omega^5}{3c^5}\text{Im}(\mathbf{p}\cdot\mathbf{t}^*) + \frac{2\omega^6}{3c^7}|\mathbf{t}|^2\right), \quad (\text{I}.1)$$

where $\omega$ is the angular frequency. The scattering cross section can be determined by $C_{sca} = 2Z_0 I / |E_0|^2$, with contributions from the irreducible dipoles and the toroidal dipole as

$$C_{sca}^{\mathbf{p}} = \frac{Z_0 \omega^4}{3\pi\varepsilon_0 c^3 |E_0|^2}|\mathbf{p}|^2, \quad C_{sca}^{\mathbf{m}} = \frac{Z_0 \omega^4}{3\pi\varepsilon_0 c^5 |E_0|^2}|\mathbf{m}|^2,$$

$$C_{sca}^{\mathbf{t}} = \frac{Z_0 \omega^6}{3\pi\varepsilon_0 c^7 |E_0|^2}|\mathbf{t}|^2, \quad C_{sca}^{\mathbf{p+t}} = \frac{Z_0 \omega^4}{3\pi\varepsilon_0 c^3 |E_0|^2}\left|\mathbf{p}+\frac{ik}{c}\mathbf{t}\right|^2, \quad (\text{I}.2)$$

where $Z_0$ is the impedance of free space and $|E_0|$ is the magnitude of the incident electric field. The term $C_{sca}^{\mathbf{p+t}}$ denotes a sum of the scattering cross sections contributed by the irreducible electric dipole only ($C_{sca}^{p}$), the toroidal dipole only ($C_{sca}^{t}$) and the interference between the two dipoles. Equation (I.2) has been used for the calculation of the results in Figs. 1(c) and 3 in the main text.

**(2) For renormalized dipoles:**

To determine the scattering cross section for the renormalized dipoles, we consider the vector spherical harmonics defined by Bohren and Huffman [3], where even and odd modes were distinguished. The corresponding scattering coefficients can be worked out using the relations:

$$A_{emn} = iE_{mn}\left[a_{m,n} + (-1)^m a_{-m,n}\right], \quad A_{omn} = -E_{mn}\left[a_{m,n} - (-1)^m a_{-m,n}\right], \quad (I.3)$$
$$B_{emn} = iE_{mn}\left[b_{m,n} + (-1)^m b_{-m,n}\right], \quad B_{omn} = -E_{mn}\left[b_{m,n} - (-1)^m b_{-m,n}\right],$$

where the subscript "e" and "o" denote even and odd, respectively, and $E_{mn} = |E_0| i^n \left[\frac{(2n+1)(n-m)!}{n(n+1)(n+m)!}\right]^{1/2}$. Note that Eq. (I.3) is applicable to the condition $m > 0$. For the condition $m = 0$, we have $A_{emn} = iE_{mn} a_{m,n}$, $B_{emn} = iE_{mn} b_{m,n}$ and $A_{omn} = B_{omn} = 0$.

The advantage to distinguish even and odd modes here is that they have one to one correspondence with the renormalized dipoles oscillating along axis directions in the Cartesian coordinate. More specifically, under the coordinate defined in the inset of Fig. 2(a) in the main text, the $A_{e11}$, $A_{o11}$ and $A_{e01}$ modes correspond to the renormalized electric dipole oscillating in the $x$-, $y$- and $z$-direction, respectively, while the $B_{e11}$, $B_{o11}$ and $B_{e01}$ modes correspond to the renormalized magnetic dipole oscillating in the $x$-, $y$- and $z$-direction, respectively.

With the help of these scattering coefficients, the scattering cross section can be evaluated by [3]

$$C_{sca} = \frac{2\pi}{k^2} \sum_{n,m}(2n+1)\left(|A_{emn}|^2 + |A_{omn}|^2 + |B_{emn}|^2 + |B_{omn}|^2\right), \quad (I.4)$$

from which we can extract the contributions from the renormalized dipoles as

$$C_{sca}^{\overline{p_x}} = \frac{6\pi}{k^2}|A_{e11}|^2, \quad C_{sca}^{\overline{p_y}} = \frac{6\pi}{k^2}|A_{o11}|^2, \quad C_{sca}^{\overline{p_z}} = \frac{6\pi}{k^2}|A_{e01}|^2,$$
$$C_{sca}^{\overline{m_x}} = \frac{6\pi}{k^2}|B_{e11}|^2, \quad C_{sca}^{\overline{m_y}} = \frac{6\pi}{k^2}|B_{o11}|^2, \quad C_{sca}^{\overline{m_z}} = \frac{6\pi}{k^2}|B_{e01}|^2. \quad (I.5)$$

Equation (I.5) has been used for the calculation of the results in Figs. 2(a) and 3 in the main text.

**II Calculation of optical forces based on field-representation**

In this part, we give a detailed description for calculating the optical force using the multipole expansion based on vector spherical functions.

The procedure starts from the expansion of the scattered fields by means of vector spherical functions. These are denoted as $\mathbf{N}_{mn}^{(J)}(k,\mathbf{r})$ and $\mathbf{M}_{mn}^{(J)}(k,\mathbf{r})$. Here, $n$ is a positive integer and $m$ is an integer ranging from $-n$ to $n$. These vector spherical functions are defined as

$$\mathbf{N}_{mn}^{(J)}(k,\mathbf{r}) = \left[\tau_{mn}(\cos\theta)\mathbf{e}_\theta + i\pi_{mn}(\cos\theta)\mathbf{e}_\phi\right]\frac{\left[\tilde{z}_n^{(J)}(kr)\right]'}{kr}\exp(im\phi)$$
$$+ \mathbf{e}_r n(n+1)P_n^m(\cos\theta)\frac{\tilde{z}_n^{(J)}(kr)}{(kr)^2}\exp(im\phi), \quad (II.1)$$
$$\mathbf{M}_{mn}^{(J)}(k,\mathbf{r}) = \left[i\pi_{mn}(\cos\theta)\mathbf{e}_\theta - \tau_{mn}(\cos\theta)\mathbf{e}_\phi\right]\frac{\tilde{z}_n^{(J)}(kr)}{kr}\exp(im\phi),$$

where the two auxiliary functions are

$$\pi_{mn}(\cos\theta) = \frac{m}{\sin\theta}P_n^m(\cos\theta), \quad \tau_{mn}(\cos\theta) = \frac{d}{d\theta}P_n^m(\cos\theta), \quad (II.2)$$

and the spherical polar coordinate related terms $r$, $\theta$ and $\phi$ are defined in the inset of Fig. 2(a), $k$ is the wave number in vacuum, $P_n^m(x)$ is the associated Legendre function of the first kind, and $\tilde{z}_n^{(J)}(x) = xz_n^{(J)}(x)$ are Ricatti Bessel functions, with

$z_n^{(J)}(x) = j_n(x), y_n(x), h_n^{(1)}(x)$, and $h_n^{(2)}(x)$ for $J$= 1, 2, 3, 4, respectively. The electric and magnetic scattered fields can then be written as an infinite series of the vector spherical functions,

$$\mathbf{E}_s = \sum_{n,m} iE_{mn}\left[a_{mn}\mathbf{N}_{mn}^{(3)}(k,\mathbf{r}) + b_{mn}\mathbf{M}_{mn}^{(3)}(k,\mathbf{r})\right],$$
$$\mathbf{H}_s = \frac{k}{\omega\mu_0}\sum_{n,m} E_{mn}\left[b_{mn}\mathbf{N}_{mn}^{(3)}(k,\mathbf{r}) + a_{mn}\mathbf{M}_{mn}^{(3)}(k,\mathbf{r})\right], \quad \text{(II.3)}$$

where $a_{mn}$ and $b_{mn}$ are the scattering coefficients for the renormalized electric multipoles and the renormalized magnetic multipoles, respectively. These scattering coefficients can be determined by, taking $a_{mn}$ for instance,

$$a_{mn} = \frac{\int_0^{2\pi}\int_0^{\pi} \mathbf{E}_s \cdot \mathbf{N}_{mn}^{(3)*}\sin\theta \, d\theta \, d\varphi}{iE_{mn}\int_0^{2\pi}\int_0^{\pi}\left|\mathbf{N}_{mn}^{(3)}\right|^2 \sin\theta \, d\theta \, d\varphi}, \quad \text{(II.4)}$$

where $E_{mn} = |E_0|i^n\left[\dfrac{(2n+1)(n-m)!}{n(n+1)(n+m)!}\right]^{1/2}$.

Meanwhile, the incident fields can be decomposed, using coefficients $p_{mn}$ and $q_{mn}$ as

$$\mathbf{E}_i = -\sum_{n,m} iE_{mn}\left[p_{mn}\mathbf{N}_{mn}^{(1)}(k,\mathbf{r}) + q_{mn}\mathbf{M}_{mn}^{(1)}(k,\mathbf{r})\right],$$
$$\mathbf{H}_i = -\frac{k}{\omega\mu_0}\sum_{n,m} E_{mn}\left[q_{mn}\mathbf{N}_{mn}^{(1)}(k,\mathbf{r}) + p_{mn}\mathbf{M}_{mn}^{(1)}(k,\mathbf{r})\right]. \quad \text{(II.5)}$$

By combining Eq. (II.3) with Eq. (II.5), we can have the explicit expressions of the electric and magnetic total fields, which can be decomposed into a sum of three parts including the incident fields ($p_{mn}$, $q_{mn}$) and the scattered fields from the renormalized electric multipoles ($a_{mn}$) and the renormalized magnetic multipoles ($b_{mn}$). The $z$-component optical force can then be derived as [4]

$$F_z = -\frac{4\pi\varepsilon_0}{k^2}|E_0|^2\sum_{n,m}\text{Re}\left(g_{mn}f_{mn}^1 + l_{mn}f_{mn}^2\right), \text{(II.6)}$$

where we have defined

$$g_{mn} = \left[\frac{n(n+2)(n-m+1)(n+m+1)}{(n+1)^2(2n+1)(2n+3)}\right]^{1/2},$$

$$l_{mn} = \frac{m}{n(n+1)},$$

$$f^1_{mn} = a_{m,n}a^*_{m,n+1} + b_{m,n}b^*_{m,n+1} - 1/2\left(a_{m,n}p^*_{m,n+1} + p_{m,n}a^*_{m,n+1} + b_{m,n}q^*_{m,n+1} + q_{m,n}b^*_{m,n+1}\right),$$

$$f^2_{mn} = a_{m,n}b^*_{m,n} - 1/2\left(p_{m,n}b^*_{m,n} + q_{m,n}a^*_{m,n}\right).$$

(II.7)

From the force expression in Eq. (II.6), we can extract the forces acting on the renormalized dipoles and the recoil forces due to dipole interference as

$$F_z^{\bar{\mathbf{p}}} = \frac{2\pi\varepsilon_0}{k^2}|E_0|^2 \sum_{n=1,m} \mathrm{Re}\left[g_{mn}a_{m,n}p^*_{m,n+1} + l_{mn}q_{m,n}a^*_{m,n}\right],$$

$$F_z^{\bar{\mathbf{m}}} = \frac{2\pi\varepsilon_0}{k^2}|E_0|^2 \sum_{n=1,m} \mathrm{Re}\left[g_{mn}b_{m,n}q^*_{m,n+1} + l_{mn}p_{m,n}b^*_{m,n}\right],$$

$$F_z^{\overline{int}} = -\frac{4\pi\varepsilon_0}{k^2}|E_0|^2 \sum_{n=1,m} \mathrm{Re}\left(l_{mn}a_{m,n}b^*_{m,n}\right).$$

(II.8)

**III Additional computational results to support the conclusions in the main text**

In this part, we give some additional results to support the conclusions in the main text.

**1. Magnetic dipole resonance in a single gold helix**

We show here that a single gold helix can support a strong magnetic dipole resonance. The gold helix is shown in Fig. S1(a) which will be used as a basic building block to construct toroidal objects in the main text. In the optical regime, the permittivity of gold can be described by the Drude model with plasma frequency $\omega_p = 2\pi \times 2.175 \times 10^{15}$ rad/s and damping frequency $\omega_c = 2\pi \times 6.5 \times 10^{12}$ rad/s [5]. Under the incidence of a linearly *x*-polarized plane wave propagating in the *z*-direction, this structure supports a magnetic dipole resonance. Figure S1(b) shows the calculated scattering cross sections using Eq. (I.2) for some components of the irreducible dipoles and the toroidal dipole in the optical regime. The *y*-component of the irreducible magnetic dipole moment ($m_y$) exhibits a stronger peak of the scattering cross section at ~66 THz, which coincides with the profile of the z-component of the optical forces acting on the irreducible electric and magnetic dipoles and the toroidal dipole, as shown in Fig. S1(c). The resonant force at ~66 THz is mainly contributed by the induced irreducible magnetic dipole moment. In particular, the effect from the toroidal dipole is completely masked by the irreducible electric and magnetic dipoles, as can be seen from Fig. S1(b), obscuring any distinction between the irreducible and the renormalized dipoles, as in most conventional structures. In other words, the results based on the field-representation are found to be graphically indiscernible from those in Figs. S1(b) and S1(c) (that is, $C_{\text{sca}}^{\bar{p}_x} \approx C_{\text{sca}}^{p_x}$, $C_{\text{sca}}^{\bar{m}_y} \approx C_{\text{sca}}^{m_y}$, $F_z^{\bar{\mathbf{p}}} \approx F_z^{\mathbf{p}}$, $F_z^{\bar{\mathbf{m}}} \approx F_z^{\mathbf{m}}$) and therefore not shown here.

**2. Toroidal force in toroidal nanostructures with different handedness**

To show the concept of the toroidal moment is useful in understanding the optical forces, we change the handedness of some helixes in Fig. 1(a) and recalculate the corresponding optical forces. The results are shown in Fig. S2, with the structure shown in the inset. We note that regardless of the handedness of the helix, the toroidal dipole resonance can always be identified and can be associated with a predominant contribution to the optical forces if we take the viewpoint that the optical force acting on an object is a consequence of the incident field interacting with charge/current distributions induced on the object.

**3. Toroidal force at microwave regime**

Our theory can also be applied to microwave regime. We consider the structure shown in the inset of Fig. S1, and change the structure parameters to inner radius $r = 0.3$ mm, outer radius $R = 1$ mm and pitch $p = 2$ mm. The resonant frequency of the toroidal dipole will then shift to microwave regime, where the conductivity of gold is

$\sigma_{Au} = 4.098 \times 10^7 \text{S/m}$. Figure S3(a) shows the calculated optical forces acting on the irreducible dipoles and the toroidal dipole. We find a strong resonant force at ~8 GHz which can be attributed to the toroidal dipole moment. This resonant force can also be completely accounted for by the renormalized electric dipole, as noted in Fig. S3(b) which shows the optical forces acting on the renormalized dipoles. In Fig. S3(a), there is a slight discrepancy between the force summation ($F_z^{\mathbf{p}} + F_z^{\mathbf{m}} + F_z^{\mathbf{t}} + F_z^{\text{int}}$) and the exact force ($F_z^{\text{exact}}$) at higher frequency range, which may be due to the relative large scatterer compared to the resonant wavelength at microwave regime. Figure S3(c) shows the scattering cross sections for the irreducible dipoles, the toroidal dipole and the renormalized dipoles, which can support the relations ($\mathbf{p} + \dfrac{ik}{c}\mathbf{t} = \overline{\mathbf{p}}, \quad \mathbf{m} = \overline{\mathbf{m}}$, for long-wavelength condition) very well, with some discrepancies at higher frequency range that can also be attributed to the relatively large scatterer. We also note that the induced dipole moments $p_y$, $t_y$ and $m_x$ are even stronger than those ($p_x$, $t_x$, $m_y$) along the direction of the corresponding incident fields. However, since the oscillating directions of these stronger dipoles are orthogonal with the corresponding incident fields, their contributions to the incident force (see Eq. (4)) are zero.

**4. Optical forces acting on split-ring-resonators (SRRs) based structure**

Our conclusion should be applicable to other systems that support toroidal dipole resonances. We consider a toroidal structure consisting of four U-shaped SRRs made of gold, as shown in the inset of Fig. S4(a), which has been studied in the literature [2]. Based on the methods introduced in the main text, we calculate the optical forces acting on the SRRs based toroidal structure. The results based on source-representation are given in Fig. S4(a), and those based on field-representation are shown in Fig. S4(b). We find the conclusion in the main text could be supported again by these results. If we view the optical force as a consequence of the interaction between the incident field and induced charge/current distributions (Fig. S4(a)), the force at the resonance at ~160 THz can be attributed to the irreducible electric dipole moment and the toroidal dipole moment. From the viewpoint of field-representation (Fig. S4(b)), this resonant force can be attributed to the renormalized electric dipole moment.

**Supplemental Figures**

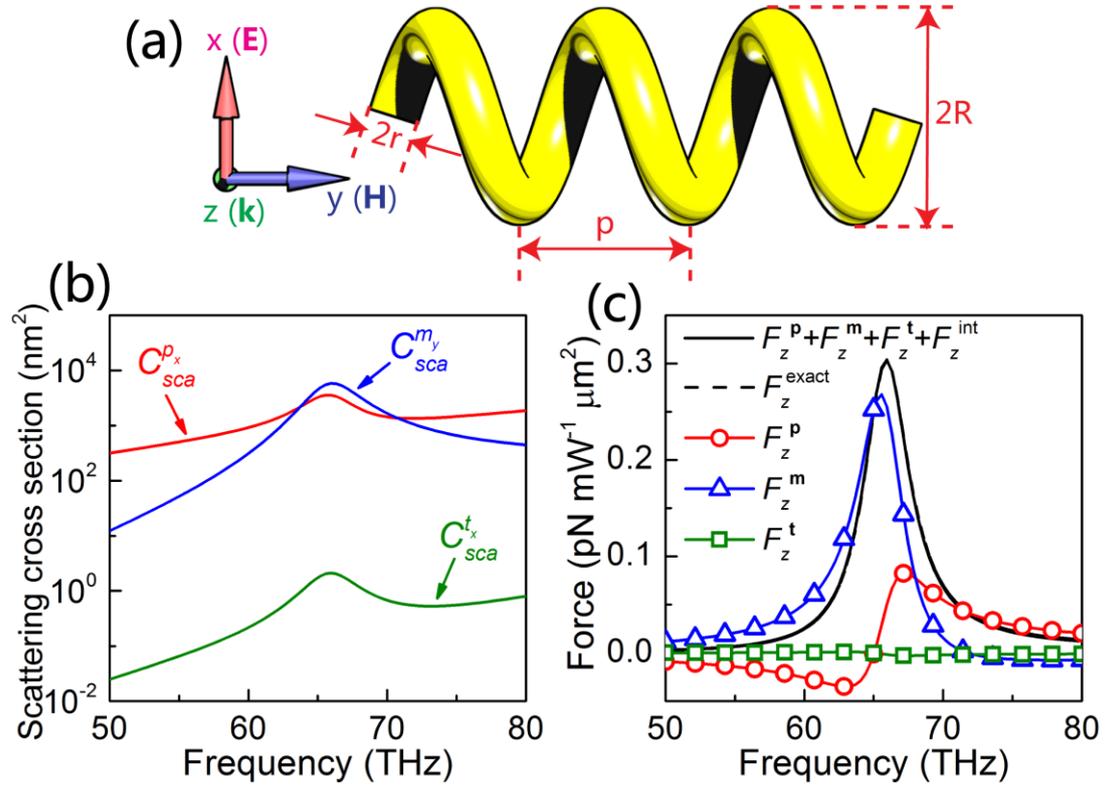

FIG. S1. (a) Schematic diagram of a gold helix that supports a magnetic dipole resonance in the optical regime. Structural parameters are inner radius $r$ = 30nm, outer radius $R$ = 100nm, and pitch $p$ = 200nm. (b) Calculated scattering cross sections and (c) $z$-component of optical forces acting on the irreducible electric and magnetic dipoles as well as the toroidal dipole in the source-representation. The incident wave is a linearly $x$-polarized plane wave propagating in the $z$-direction. The force $F_z^{exact}$ is calculated according to Eq. (1) using numerically obtained scattered fields.

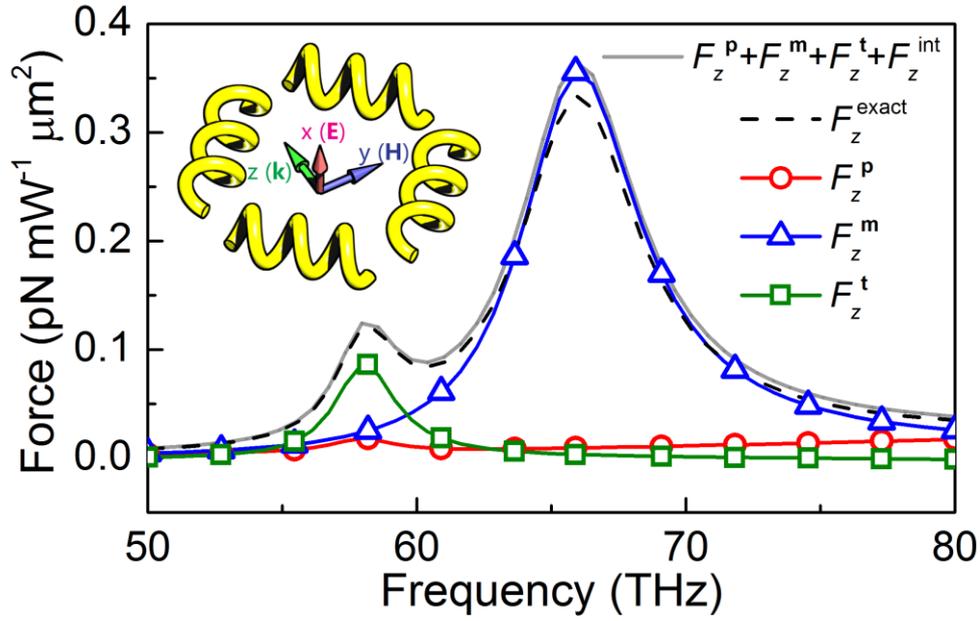

FIG. S2. Calculated *z*-component of the optical forces in the optical regime acting on the irreducible dipoles and the toroidal dipole based on source-representation. The inset of the figure illustrates the studied toroidal structure with different selection of the helical handedness compared to the one in Fig. 1(a). The arrows show the direction of the *k*-vector of the incident plane wave and its polarization.

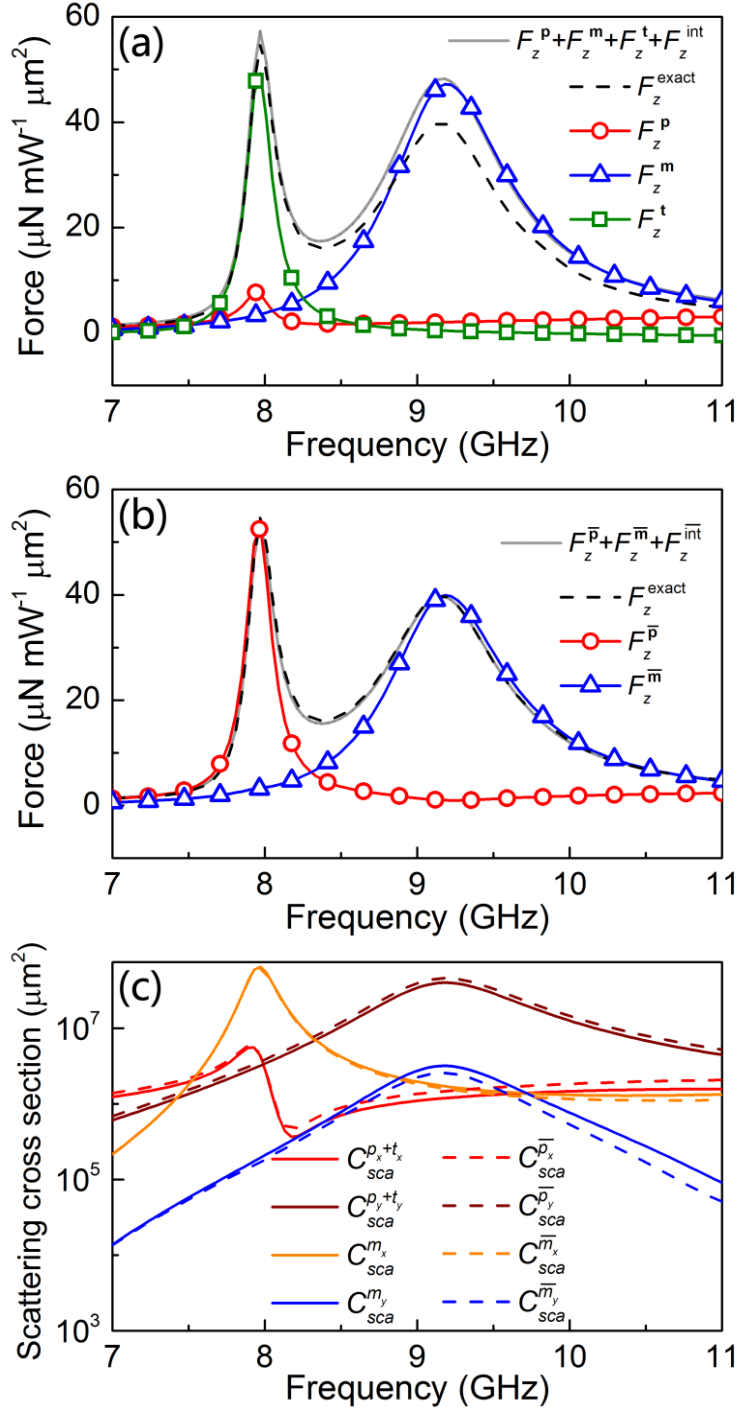

FIG. S3. Calculated $z$-component of the optical forces in the microwave regime acting on the toroidal structure (see the inset of Fig. S1) based on (a) source-representation and (b) field-representation. The parameters are inner radius $r = 0.3$ mm, outer radius $R = 1$ mm, and pitch $p = 2$ mm. (c) Calculated scattering cross sections for the irreducible dipoles, the toroidal dipole, and the renormalized dipoles.

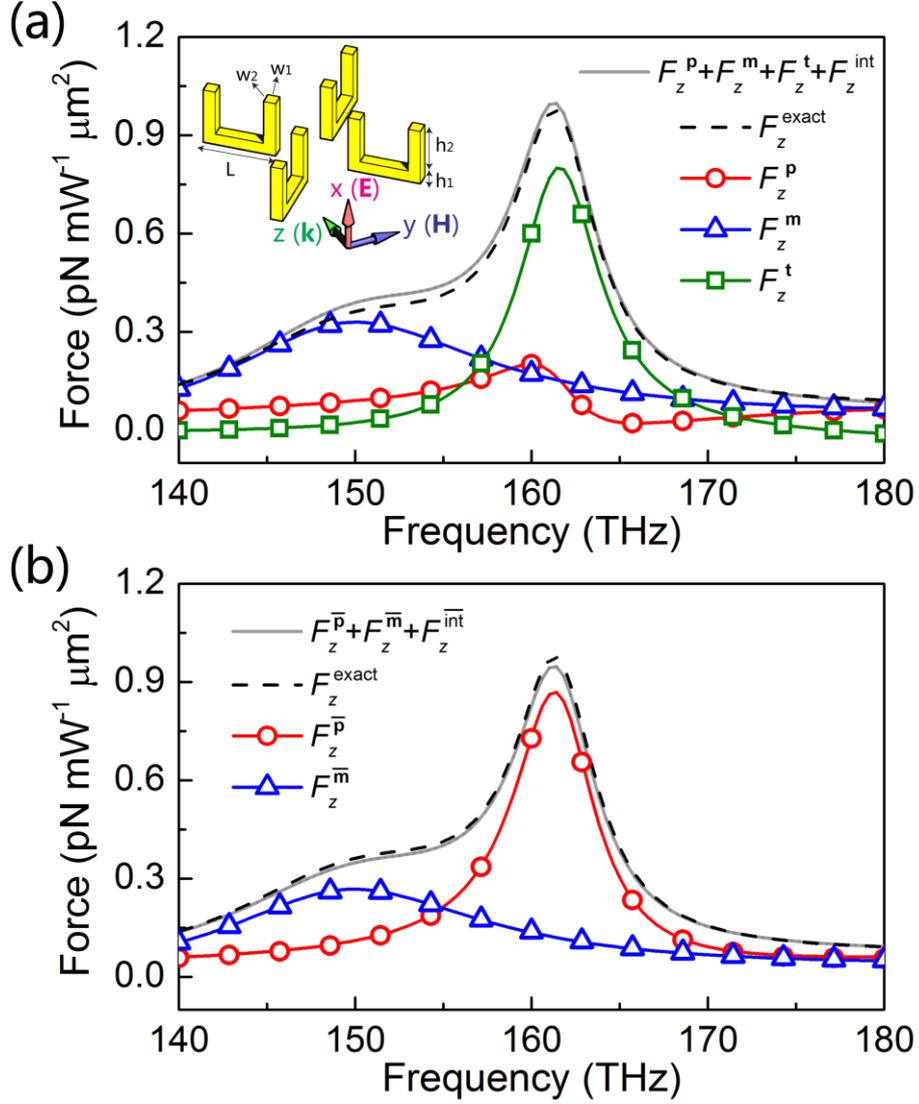

FIG. S4. Calculated *z*-component of the optical forces acting on the U-shaped SRRs based toroidal structure based on (a) source-representation and (b) field-representation. The parameters of each SRR are $L = 300$ nm, $h_1 = w_1 = w_2 = 50$ nm, $h_2 = 200$ nm, and the distance between the structure center and the center of each SRR is 300 nm. The structure is shown in the inset of the top panel. The arrows show the direction of the *k*-vector of the incident plane wave and its polarization. The resonant optical force at slightly above 160 THz can be attributed mainly to a toroidal dipole moment (green line, upper panel) or to renormalized electric dipole moment (red line, lower panel) depending on how we choose to do the multipole expansion.